\documentclass[10pt,a4paper]{article}
\usepackage[latin1]{inputenc}
\usepackage{amsmath}
\usepackage{amsfonts}
\usepackage{amssymb}
\usepackage{graphicx}

\author{Joshua I. James\textsuperscript{1}, Yunsik Jake Jang\textsuperscript{2}\\
joshua@cybercrimetech.com, ccismem@gmail.com\\
\\
\textsuperscript{1}Digital Forensic Investigation Research Group\\
University College Dublin\\
Belfield, Dublin 4, IE\\
\\
\textsuperscript{2}International Cybercrime Research Center\\
Korean National Police University\\
Yongin-si, South Korea}
\title{An Assessment Model for Cybercrime Investigation Capacity}
\date{}
\begin{document}
\maketitle

\begin{abstract}
Digital technologies are constantly changing, and with it criminals are finding new ways to abuse these technologies. Cybercrime investigators, then, must also keep their skills and knowledge up to date. This work proposes a holistic training development model -- specifically focused on cybercrime investigation -- that is based on improving investigator capability while also considering the capacity of the investigator or unit. Along with a training development model, a cybercrime investigation capacity assessment framework is given for attempting to measure capacity throughout the education process. First, a training development model is proposed that focuses on the expansion of investigation capability as well as capacity of investigators and units. Next, a capacity assessment model is given to evaluate the effectiveness of the training program. A description of how the proposed model is being applied to the development of training programs for cybercrime investigators in developing countries will then be given, as well as already observed challenges. Finally, concluding remarks as well as proposed future work is discussed.
\end{abstract}


\section{Introduction}
Continuous development of information communication technology and large scale proliferation is leading to an increase in victims of digital crime, as well as tools and evidence in criminal investigations [1]. Many crimes, even traditionally non-digital crimes such as murder, now normally have some sort of digital component [2,3]. Even so much so that some law enforcement agencies claim that there are rarely cases that do not involve some sort of digital device [4]. Because of this, the ability of law enforcement to handle electronic evidence is vital during entire investigative procedure. A survey recently conducted by the United Nations (U.N.) reported that more than 90\% of respondent countries have at least some capability to conduct digital investigations [5].

However, the U.N. report also showed a global need for expansion of digital investigation capability. Key areas of concern were with the investigation capabilities in less developed countries, as well as the capability of organizations to conduct international investigations. This mirrors Huber [6], who claimed that ``digital forensics is very much an interstate and international issue''. General cybercrime investigation, computer forensics and digital evidence handling were the most commonly cited areas that require more development and assistance. Currently, many public and private organizations around the world are investing in the infrastructure of developing countries. However, international assistance in regards to digital crime investigation has been slower that other sectors, such as information technology [7].

Many countries, developing and developed alike, are beginning to invest in the capability of their investigators to investigate digital crimes. This has caused a rapid growth in the number of digital crime investigation programs offered at training institutions and universities around the world. These programs are targeted at improving an investigator's \textit{capability} in dealing with certain types of digital crime; however, improvement of capability does not necessarily improve investigation \textit{capacity}.

Hekim, Gul et al. [8] previously examined investigation capacity in law enforcement to assess whether information technology improves investigation capacity. In their work, they define investigation capacity as the case clearance rate. James and Jang [9], however, claim that the case clearance rate alone does not adequately represent an investigator or unit's investigation capacity. They instead define investigation capacity as case clearance rates as related to time, where:\\

\begin{center}
$capacity(T)=((\frac{cases\ closed}{average\ investigators})+((\frac{cases\ closed}{average\ investigators}) \cdot down\ time)) \cdot (average\ investigators)$
\end{center}

\noindent where:
\begin{itemize}
\item $T$ is the time span for which capacity should be measured
\end{itemize}

Capability and capacity are related terms that have to do with quality. In investigations a highly capable investigator may be able to conduct quality investigations unless there are more investigation requests than he or she has the capacity to deal with at a time, at which point quality may be reduced. Many investigators have to cope with a lack of resources [10], meaning that they are normally working at -- if not over -- capacity. Because of this, when attempting to educate investigators both expansion of capability and capacity should be considered and measured.

\subsection{Contribution and Structure}
This work proposes a holistic training development model -- specifically focused on cybercrime investigation -- that is based on improving investigator capability while also considering the capacity of the investigator or unit. Along with a training development model, a cybercrime investigation capacity assessment framework is given for attempting to measure capacity throughout the education process.

First, a training development model is proposed that focuses on the expansion of investigation capability as well as capacity of investigators and units. Next, a capacity assessment model is given to evaluate the effectiveness of the training program. A description of how the proposed model is being applied to the development of training programs for cybercrime investigators in developing countries will then be given, as well as already observed challenges. Finally, concluding remarks as well as proposed future work will be discussed.

\section{Building Investigation Capability}
The growing number and complexity of digital devices related to criminal acts means that digital investigators must attempt to increase their knowledge of a broad range of technologies, as well as increase the pace at which such devices can be analyzed [11]. There are a number of challenges that digital investigators face, and education and training is a primary challenge for many investigation units today [7,10]. Not only is receiving training sometimes a challenge, but the way training is organized and delivered can also cause issues. Further, in many training programs the content may be such that investigators don't -- or can't -- implement trained material in practical investigations. Such a situation could be the product of a training program that did not meet the needs of the trainees; however, without post-training assessment there is no way to know whether skills and knowledge are being transferred to practice. The Korean National Police, for example, are active contributors for building investigation capability and capacity with developing countries; however, few development programs have adequately evaluated the effectiveness of such initiatives.

With many organizations, evaluation may be at the end of a training session and perhaps award the trainee a simple qualification upon completion. However, short term trainings do not necessarily lead to long-term knowledge gain, even if a trainee does well on an examination immediately after the course [7]. Also, because no further evaluation is normally done, training organizations have no way of assessing if knowledge or skills have been practically transferred.

For the issues of generally targeting the training to the needs of the trainees, and ensuring that knowledge and skills are being practically transferred for the improvement of capability and capacity, this work proposes a holistic training development methodology to increase cybercrime investigation capability and capacity.

\begin{figure}
  \centering
  \includegraphics[width=0.9\textwidth]{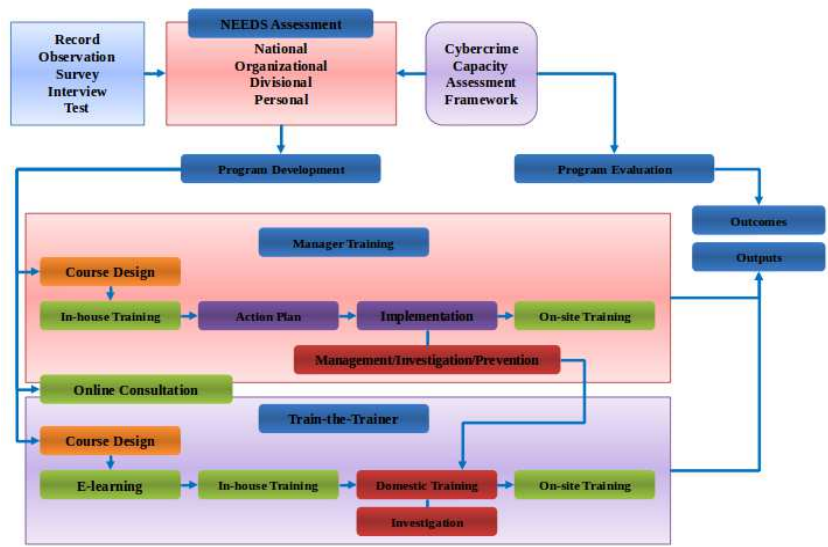}
  \caption{Cybercrime investigation capacity expansion framework with a focus on human resource training}
  \label{fig:PDMdiagram}
\end{figure}

As shown in Figure \ref{fig:PDMdiagram}, the method begins with an extensive assessment of the organization, attempting to determine what their goals, processes and current knowledge are. During the observation phase, a training group should gain understanding of the current state of knowledge and skills of the organization to be trained as well as how the organization can/will implement new knowledge. In our experience, it is common for the trainers to find that organizations to be trained are either already knowledgeable about many of the areas, or that the organization has no skills in the area and may require remediation. It is relatively uncommon to have a group, even within the same organization, in which all participants have the same level of knowledge and skills in a particular area.

Once a better understanding of how the organization functions has been established, the needs of the organization should be determined. This stage should be approached with some skepticism as many organizations tend to have difficulty differentiating between wants and needs, and in some cases are simply ignorant of their actual needs. Based on knowledge gained in the observation phase, paired with lists of the organization's needs, a training developer should attempt to assess identified needs in the context of the organization's goals and current practices. Many organizations, for example, often identify `more equipment' as a need to expand capacity. Capacity, however, can be expanded in many ways, such is more efficient processes. The training developer, then, should attempt to determine major problem areas at the national, organizational, divisional and personal levels, and evaluate what is needed to make a lasting positive impact on the organization. Again considering more equipment as a example, additional equipment may make up for inefficient processes in the short-term, but will exacerbate inefficient processes as capacity again reaches its peak.

\subsection{Program Development}
Once needs of the organization have been established, program development can begin. There are many course development strategies, but we tend to prefer the Successive Approximation Model (SAM). SAM is essentially a breadth first approach to curricula design, where the design is iterative -- from general to specific -- for approximately three iterations [12]. SAM is composed of three specific phases: preparation, design and development.
\begin{itemize}
\item Preparation
\begin{itemize}
\item Information gathering
\begin{itemize}
\item Background information
\item Needs analysis
\end{itemize}
\end{itemize}
\item Interactive design
\begin{itemize}
\item Iterate through the design, prototype and review phases
\item ``Savvy start''
\item Project planning
\item Additional design
\end{itemize}
\item Interactive development
\begin{itemize}
\item Iterate through development, implementation and evaluation
\item Design proof
\item Alpha Version
\item Beta Version
\item Final `Gold' Version
\end{itemize}
\end{itemize}

SAM is general and intuitive enough to be easily applied to many different areas. Iteration in the development and design phases means that the developer can be more abstract and focus in once they have more information and feedback, rather than picking a design and sticking strictly with it. Further, iterations of SAM include an evaluation phase ensuring quality standards are maintained. In this case, SAM is used for course design for the manager and train-the-trainer programs.

\subsubsection{Management Training}
As discussed, many training and education initiatives focus on the expansion of capabilities of investigators. However, to improve capability and capacity, management must also be involved. In the proposed model, management receives in-house training designed to suit the organization's identified needs. Such training could focus on improving work flows, data management, evidence intake, etc. as can be seen by a work flow process analysis in [13].

Once initial in-house training has been given, management is expected to form action plans that begin to apply what they have learned. Created action plans should be practically implementable in their organization. Implementation of new knowledge is assessed and supplemented via practical 'management' on-site training. On-site training allows the trainer to better understand how -- or if -- the trainees are attempting to apply what was learned to improve processes in the organization. This also allows the trainer to help point out potential areas of improvement, and answer practical implementation questions with the organization as an example.

The outputs of management training are crime management, investigation and prevention strategies that are beginning to be implemented. This includes management's, possibly newly created, goals for investigator training and investigation needs.

\subsubsection{Train-the-Trainer}
Training for investigators is developed in parallel with manager training, again using SAM. The first content to be created for investigators is the e-learning component. E-learning allows busy organizations to complete some training at a time that is convenient for them. It also allows trainers to assess trainees' current level of knowledge, and provide remedial education, if necessary. This will allow all participants to have a similar starting baseline knowledge, which is missing in many training programs.

Next, in-house training for investigators is developed around the previously identified needs, using the same themes as the management training. Once, skills and knowledge for investigators have been developed, and practical feedback has been given from the management training, then investigators undergo domestic training that feeds directly into their investigation process. These processes continue, and are supplemented with on-site training.

\subsection{Online Consultation}
Oftentimes when organizations receive training and education, trainees may make informal social network connections that may help in the application of learned skills and knowledge. However, this support may not be responsive, consistent, or the best suited for helping organizations. For this reason, online consultation with experts associated with the training organization is provided in this model. In this case, if trainees have questions, they can consult with experts in the knowledge domain rather than other practitioners that may also be learning these processes. With such support, the trainees may be more confident about attempting to develop and implement better investigation processes and techniques.

\section{Cybercrime Investigation Capacity Assessment Framework}
The cybercrime investigation capacity assessment framework includes observation and assessment throughout the training program. Observation of the organization is conducted before and after each training initiative for the whole organization.

The observation phase is composed of a survey of the organization to determine the basic level of skills and knowledge for the organization. This survey, along with knowledge about organization processes determined during the first observation phase give the training organization a baseline with witch to compare skills and knowledge development.

After a training session has been delivered -- either in-house or on-site -- an assessment of the trainees is given to attempt to evaluate their short term understanding of the objectives. From three to six months after training has been received, the survey will again be administered to assess capability, and an assessment of changed processes will be conducted to evaluate efficiency improvements. Finally, the previously described capacity measurement model [9] will be used to assess improvements in capacity over time.

By using a qualitative survey, an observation of process changes, and a quantitative approach to capacity assessment, greater understanding of how the organization's process has change, and what the effects of this change are in terms of capacity can be determined.

\section{Application to Cybercrime Training Development}
This process has been developed in response to shortcomings observed in current cybercrime investigation training initiatives, especially in the area of assessment. The overall process is currently being applied to three (3) year cybercrime investigation capacity building programs for 5 Asian countries. In many of these countries, such a project will be the first attempt to measure the countries' cybercrime investigation capacity, and evaluate an organization's investigation process in relation with training. Such an initiative, however, is only a first effort towards measuring the effects of training on cybercrime investigators. Therefore the model, although currently general, is expected to be modified with experience. This is especially true from the cybercrime investigation capacity assessment framework as better metrics for metrics are identified.

\section{Conclusions and Future Work}
This work proposed a holistic training development model -- specifically focused on cybercrime investigation -- that is based on improving investigator capability while also considering the capacity of the investigator or unit. Along with a training development model, a cybercrime investigation capacity assessment framework was given for attempting to assess investigation capability and capacity throughout the education process.

Such a model is proposed to a attempt to improve pre and post training assessment, as well as allowing training developers more information about the organization that will receive the training so such training can be better targeted, or remediation can take place before training begins if necessary.

As discussed previously, this development and assessment model is currently being applied to training programs with five Asian countries. Ideally this model will allow training developers to understand the individual needs of each country, but will result in a relatively generalized program that meets the needs of each group. Future work will attempt to assess the weaknesses of each step of the model, especially with metrics used in the capacity assessment framework. 

\section*{Bibliography}
\begin{enumerate}
\item Casey, E. (2004). Digital evidence and computer crime : forensic science, computers and the Internet (2nd ed., pp. xviii, 690p.). Amsterdam, London: Elsevier Academic.
\item RTE. (2010). Text message evidence in murder trial. RTE News. Retrieved from http://www.rte.ie/news/2010/0423/drimnagh.html
\item Nguyen, L. (2012). Tori Stafford trial: Cellphone record shows gap during abduction, murder. Postmedia News. Retrieved from http://www.canada.com\\/life/Tori+Stafford+trial+Cellphone+record+shows+during+abduction+\\murder/6486178/story.html
\item Philipp, J. (2013). Nearly Every NYC Crime Involves Cyber, Says Manhattan DA. The Epoch Times. Retrieved from http://www.theepochtimes.com\\/n2/united-states/nearly-every-nyc-crime-involves-cyber-says-manhattan-da-355692.html
\item (2013) ``Comprehensive Study on Cybercrime''. United Nations Office on Drugs and Crime. http://www.unodc.org/documents/commissions\\/CCPCJ\_session22/13-80699\_Ebook\_2013\_study\_CRP5.pdf
\item Huber, E. (2010). Certification, Licensing, and Accreditation in Digital Forensics. A Fistful of Dongles. Retrieved from http://ericjhuber.blogspot.com\\/2010/11/certification-licensing-and.html
\item James, J. I., \& Gladyshev, P. (2013). Challenges with Automation in Digital Forensic Investigations, 17. Computers and Society. Retrieved from http://arxiv.org/abs/1303.4498
\item Hekim, H., Gul, S., \& Akcam, B. (2013). Police use of information technologies in criminal investigations. European Scientific Journal, 9(4), 221-240. Retrieved from http://eujournal.org/index.php/esj/article/view/778
\item James, J. I., \& Jang, Y. J. (2013). Measuring digital crime investigation capacity to guide international crime prevention strategies. Unpublished paper in the 7th International Symposium on Digital Forensics and Information Security. 
\item Gogolin, G. (2010). The Digital Crime Tsunami. Digital Investigation, 7(1-2), 3-8. doi:10.1016/j.diin.2010.07.001
\item Casey, E., Ferraro, M., \& Nguyen, L. (2009). Investigation Delayed Is Justice Denied: Proposals for Expediting Forensic Examinations of Digital Evidence. Journal of forensic sciences, 54(6), 1353?1364. doi:10.1111/j.1556-4029.2009.01150.x
\item Allen, M. W. (2003). Michael Allen's guide to e-learning: Building interactive, fun, and effective learning programs for any company. Wiley.
\item James, J. I., \& Gladyshev, P. (2013). A survey of digital forensic investigator decision processes and measurement of decisions based on enhanced preview. Digital Investigation, 1-10. doi:10.1016/j.diin.2013.04.005
\end{enumerate}

\end{document}